\documentclass[aps,prb,10pt,twocolumn,floatfix,superscriptaddress,showpacs,numerical,footinbib,longbibliography]{revtex4-1}

\usepackage{graphicx}
\usepackage{amsmath}
\usepackage{amssymb}
\usepackage{mathdots}
\usepackage{bbold}
\usepackage[pdftex]{color}
\usepackage[%
  colorlinks=true,
  urlcolor=blue,
  linkcolor=blue,
  citecolor=blue
]{hyperref}

\newcommand{\e}{\mathrm{e}}
\newcommand{\im}{\mathrm{i}}
\newcommand{\di}{\mathrm{d}}
\newcommand{\ket}[1]{| #1 \rangle}
\newcommand{\bra}[1]{\langle#1 |}

\begin{document}

\title{Nodal-line semimetals from Weyl superlattices}

\author{Jan Behrends}
\affiliation{Max-Planck-Institut f\"ur Physik komplexer Systeme, 01187 Dresden, Germany}
\author{Jun-Won \surname{Rhim}}
\affiliation{Max-Planck-Institut f\"ur Physik komplexer Systeme, 01187 Dresden, Germany}
\author{Shang Liu}
\affiliation{Department of Physics, Harvard University, Cambridge, MA 02138, USA}
\author{Adolfo G. Grushin}
\affiliation{Department of Physics, University of California, Berkeley, California 94720, USA}
\affiliation{Institut N{\'e}el, CNRS and Universit{\'e} Grenoble Alpes, F-38042 Grenoble, France}
\author{Jens H.\ Bardarson}
\affiliation{Max-Planck-Institut f\"ur Physik komplexer Systeme, 01187 Dresden, Germany}
\affiliation{Department of Physics, KTH Royal Institute of Technology, Stockholm, SE-106 91 Sweden}

\begin{abstract}
The existence and topological classification of lower-dimensional Fermi surfaces is often tied to the crystal symmetries of the underlying lattice systems.
Artificially engineered lattices, such as heterostructures and other superlattices, provide promising avenues to realize desired crystal symmetries that protect lower-dimensional Fermi surface, such as nodal lines.
In this work, we investigate a Weyl semimetal subjected to spatially periodic onsite potential, giving rise to several phases, including a nodal-line semimetal phase.
In contrast to proposals that purely focus on lattice symmetries, the emergence of the nodal line in this setup does not require small spin-orbit coupling, but rather relies on its presence.
We show that the stability of the nodal line is understood from reflection symmetry and a combination of a fractional lattice translation and charge-conjugation symmetry.
Depending on the choice of parameters, this model exhibits drumhead surface states that are exponentially localized at the surface, or weakly localized surface states that decay into the bulk at all energies.
\end{abstract}

\maketitle

\section{Introduction}

Symmetries play a crucial role in the realization of lower-dimensional Fermi surfaces: while Dirac semimetals with degenerate Fermi points require time-reversal in combination with crystal symmetries, such as reflection or rotational invariance, Weyl semimetals are stable even in absence of these\cite{Herring:1937do}.
However, Weyl nodes may be gapped by coupling two nodes of opposite chirality\cite{Turner:2013cj}.
Although isolated Weyl nodes are generally stable towards small disorder\cite{Sbierski:2014bo}, scattering between different Weyl nodes couples them and may open up a gap in the spectrum by annihilating the Weyl nodes.
Alternatively, we demonstrate that Weyl nodes can couple such that a new phase arises: a nodal-line semimetal with a one-dimensional Fermi surface\cite{Burkov:2011ek,Chen:2015,Kim:2015ie,Zeng2015,Yu:2015gc,Yamakage:2015im,Xie:2015hf,Bian:2016dt,Okamoto:2016,Phillips:2014,Hirayama:2017,Hu:2016zr,Huang:2016al,Mullen:2015,Xu:2017eb}.

Nodal-line semimetals exhibit surface bands at a limited range of momenta\cite{Chan:2016ho}.
These bands may serve as a basis for correlated physics in the presence of interactions\cite{Kopnin:2011jk}.
In the bulk, the implications of the nodal line include a sharply peaked magnetic susceptibility at zero energy\cite{Koshino:2016fb,Mikitik:2016}, three-dimensional integer quantum Hall effect\cite{Rhim:2015}, and intriguing transport and density response properties\cite{Burkov:2011ek,Carbotte:2016,Rhim:2016,Mukherjee:2017,Wang:2017x,Emmanouilidou:2017,Syzranov:2016x,Ahn:2017x,Matusiak:2017,Yan:2016ee}.
However, nodal-line phases generically do not survive the inclusion of spin-orbit coupling\cite{Weng:2015ec,Xu:2015ky,Weng:2015ke,Kim:2015ie,Xie:2015hf,Zeng2015,Yu:2015gc,Yamakage:2015im}, which lifts the nodal degeneracy leading to isolated Weyl points.
It is thus desirable to conceive realizations of nodal-line semimetals that do not rely on a small or vanishing spin-orbit coupling, as already suggested in previous proposals\cite{Fang:2015gt,Carter:2012ec,Li2017} that require time-reversal and inversion symmetry, or nonsymmorphic symmetries\cite{Young:2015co,Wieder:2016ha}.

In this work, we show that a Weyl semimetal subjected to a spatially periodic modulation of the onsite potential can undergo a transition to various phases, including a nodal-line semimetal.
Since spin-orbit coupling is usually a requirement to have a Weyl phase\cite{Huang:2015ig,Weng:2015ec}, the nodal-line semimetal also relies on it, contrary to other proposals that require small spin-orbit coupling\cite{Weng:2015ke,Kim:2015ie,Xie:2015hf,Zeng2015,Yu:2015gc,Yamakage:2015im}.
We present the topological classification of the nodal line, showing that its stability relies on reflection symmetry\cite{Chiu:2014fi} and a combination of a fractional lattice translation and charge-conjugation symmetry.

The nodal-line semimetal phase that we predict exhibits surface states that are not pinned to zero energy, similar to previously studied models with drumhead surface states\cite{Chan:2016ho,Bian:2016dt,Xu:2017eb}.
The extra charge accumulation due to states at the surface is tied to the intercellular Zak phase\cite{Rhim:2017bg}.
This implies that realizations without surface states exponentially localized to the boundary are possible.
Nodal-line semimetals without surface states at low energies enable the direct study of the bulk properties of the nodal line.

Periodic superlattices can be implemented both in solid-state\cite{Bouvier2011,Mogi:2017fi,Belopolski:2017cp} and synthetic systems\cite{Roati:2008hi,Deissler:2010hl,Neff:05}.
Multilayer heterostructures can effectively realize superlattices for  Weyl fermions in crystalline solids.
A Weyl phase realized on an optical lattice~\cite{Dubcek:2015jt} can be supplemented with a superlattice~\cite{Wang:2017ib} to obtain the nodal-line semimetal phase proposed here.
Since Weyl fermions are realized in photonic crystals~\cite{Lu:2013ch,Lu:2015by,Wang:2016kp}, and superlattice structures have been engineered to observe Brillouin zone folding effects~\cite{Neff:05}, these systems may also serve as a natural platform for the phenomena we study.

The paper is organized as follows.
After introducing the superlattice model in Sec.~\ref{sec:model}, we diagonalize the full Hamiltonian to obtain its band structure, revealing a nodal-line semimetal.
The emergence of the nodal line is understood from the derivation of a low-energy effective theory that we discuss in Sec.~\ref{sec:low_energy}.
To predict the stability of the nodal line we supplement the low-energy-theory with a symmetry classification based on reflection symmetry and a combination of a fractional lattice translation and charge-conjugation symmetry in Sec.~\ref{sec:symmetries}.
The circumstances under which the model exhibits drumhead surface states are understood via the intercellular Zak phase and explained in Sec.~\ref{sec:surface_states}.
The existence of the nodal line is in all cases protected by at least one symmetry, and therefore a wave-vector mismatch between the superlattice and the Weyl semimetal does not immediately open a gap, as shown in Sec.~\ref{sec:stability}.
Our results do not hinge on a specific lattice model of a Weyl semimetal; in Sec.~\ref{sec:tr_symmetric} we demonstrate that a nodal line also arises in a superlattice of a time-reversal invariant Weyl semimetal.

\section{Model: Weyl semimetal on a superlattice}
\label{sec:model}

We start from the time-reversal-breaking two-band lattice Hamiltonian\cite{Yang:2011im}
\begin{equation}
 \mathcal{H}_0 (\mathbf{k}) =  v\,\left( \sin k_x \,\sigma_x + \sin k_y \,\sigma_y \right) + M_\mathbf{k}\,\sigma_z,
\label{eq:two_band_model}
\end{equation}
where $M_\mathbf{k} = t\,(2- \cos k_x - \cos k_y) + v \left(\cos k_z -m \right)$ and the lattice constant is set to $a=1$.
For certain values of $m$, e.g.,\ $-1<m<5$ at $v=t$, the model describes a Weyl semimetal\cite{Chen:2015he}.
The Hamiltonian obeys a charge-conjugation symmetry, which may be broken in higher-energy bands in more realistic systems. In the course of this work, we consider this symmetry to be fulfilled.
Here, we focus on $0<m<1$, when there is one pair of Weyl nodes at $\left(0,0, \pm \arccos m \right)$.
This Hamiltonian is perturbed by the periodic potential
\begin{equation}
 U (\mathbf{r})
	= 2\,\sum_{\mu=\lbrace 0,x,y,z \rbrace} \,u_\mu\,\cos \left( \mathbf{r}\cdot \mathbf{K} - \theta_\mu \right)\,\sigma_\mu,
 \label{eq:perturbation}
\end{equation}
where $\sigma_\mu=(\sigma_0,\boldsymbol{\sigma})$, $\sigma_0$ is the $2\times2$ identity matrix, $\boldsymbol{\sigma}$ is the vector of Pauli matrices, and $\mu=0,x,y,z$.
Depending on the physical realization of this low-energy Hamiltonian, $\sigma_\mu$ may act in spin or orbital space, or, when realized on an optical lattice, in sublattice space\cite{Dubcek:2015jt}.
The angles $\theta_\mu$ denote a shift of each component of the periodic potential towards the original lattice.

In the interest of clarity, we make two provisional simplifying assumptions: 
first, we assume that $\mathbf{K}$ is commensurate with a reciprocal-lattice vector in the $z$-direction, which sets $\mathbf{K} = (2\,\pi/n)\,\mathbf{e}_z$ for a folding degree $n \in \mathbb{N}$ and the unit vector $\mathbf{e}_z$.
Second, the vector $\mathbf{K}$ is chosen to match the wave vector connecting the two Weyl nodes, $\mathbf{K} = 2\,\arccos m\,\mathbf{e}_z$, thereby restricting our discussion to the specific values of $m = \cos \pi/n$.
We postpone discussing the consequences of relaxing these two assumptions to Sec.~\ref{sec:stability}.

\begin{figure}
 \includegraphics[width=\linewidth]{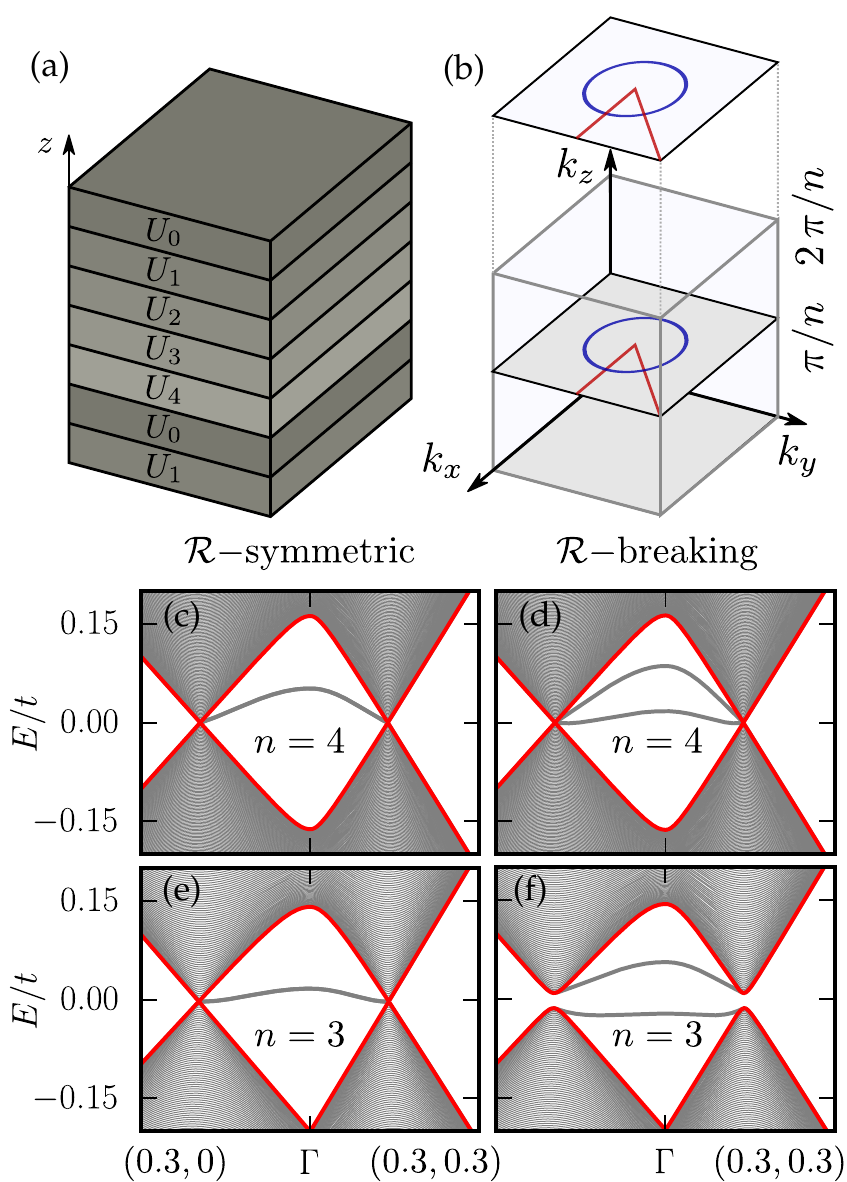}
 \caption{(a) Stacking of layers of Weyl semimetals subjected to different potentials $U_0$ with a folding degree $n=5$, realizing the superlattice discussed in this work.
 (b) Brillouin zone and surface Brillouin zone projected on the $(001)$ surface with the position of the nodal line that may emerge in the $k_x$-$k_y$ plane at $k_z = \pi/n$, depending on the choice of parameters.
 The dispersion of the full Hamiltonian, Eq.~\eqref{eq:hamiltonian}, along the two dark red lines in the bulk and surface Brillouin zone is plotted in (c) to (f), with $v=t, u_0 = 0.2\,t$ and $u_{x,y,z} = 0$ for different angles $\theta_0$.
 The red lines show the dispersion at $k_z = \pi/n$ and the gray lines the dispersion for open boundary conditions in the $z$-direction with 256 sites, including surface states.
 The different plots show (c) folding degree $n=4$, obeying reflection symmetry with $\theta_0 = 3/4\,\pi$, (d) $n=4$, breaking reflection symmetry with $\theta_0 = 3/4\,\pi + 0.15$, (e) $n=3$, obeying reflection symmetry with $\theta_0 = 2/3\,\pi$, (f) $n=3$, breaking reflection symmetry with $\theta_0 =2/3\,\pi +0.15$.
 }
 \label{fig:dispersion}
\end{figure}

Such choice of $\mathbf{K}$ enlarges the unit cell by the folding degree $n$ in the $z$-direction, as shown in Fig.~\ref{fig:dispersion}~(a).
The full Hamiltonian can be written in a form representing the larger unit cell
\begin{equation}
 \mathcal{H}_{n\mathbf{k}} = 
 \begin{pmatrix}
   h_\mathbf{k} + U_0 & & & \mathrm{h.c.} \\
  \tfrac{v}{2}\, \sigma_z & h_\mathbf{k} + U_1 &  & \\
  & \ddots & \ddots &  \\
  \tfrac{v}{2}\,\e^{-\im\,n\, k_z}\,\sigma_z & & \tfrac{v}{2} \,\sigma_z & h_\mathbf{k} + U_{n-1}
 \end{pmatrix}.
\label{eq:hamiltonian}
\end{equation}
There is a gauge freedom in choosing the phases of the basis functions for the sites that constitute the superlattice.
Our gauge choice ensures that the Hamiltonian is invariant under a shift by a reciprocal-lattice vector.
The $j$-th diagonal element of the Hamiltonian is $h_\mathbf{k}+U_j$ with
\begin{equation}
 h_\mathbf{k}	=  \mathcal{H}_0 (\mathbf{k}) - v\,\cos k_z\,\sigma_z  ,~
 U_j		= U \left( \mathbf{r} = j\,\mathbf{e}_z \right) .
\end{equation}

To gain further intuition of the nature of the perturbation, one can express Eq.~\eqref{eq:hamiltonian} as $n$ copies of a Weyl Hamiltonian at different momenta, coupled by the superlattice perturbation~\eqref{eq:perturbation}.
To this end, we rotate the Hamiltonian using the unitary transformation $\mathcal{V} = \mathcal{V}_0 \otimes \sigma_0$ with elements
\begin{equation}
 \left( \mathcal{V}_0 \right)_{jl} = \frac{1}{\sqrt{n}} \,\exp \left[ -\im\,l \left( k_z + \frac{2\,\pi\,j}{n} \right) \right],
\end{equation}
to obtain $\tilde{\mathcal{H}}_{n \mathbf{k}} = \mathcal{V}\,\mathcal{H}_{n\,\mathbf{k}} \,\mathcal{V}^\dagger$,
 \begin{equation}
 \tilde{\mathcal{H}}_{n \mathbf{k}} =
 \begin{pmatrix}
   \mathcal{H}_0 (\mathbf{k})	& U_+ & & U_- \\
   U_- & \mathcal{H}_0 (\mathbf{k} + \tfrac{2\pi \mathbf{e}_z }{n} ) & U_+ & \\
   & & \ddots & \\
   U_+ & & U_- & \mathcal{H}_0 (\mathbf{k} +  \tfrac{2 \pi(n-1) \mathbf{e}_z }{n}  )
 \end{pmatrix}
 \label{eq:hamiltonian_alternative}
\end{equation}
with the Weyl Hamiltonian $\mathcal{H}_0 (\mathbf{k})$ and the perturbation $U_\pm = \sum_\mu u_\mu\,\e^{\pm\im\,\theta_\mu}\,\sigma_\mu$.
Thus, the periodic perturbation couples Weyl Hamiltonians evaluated at momenta that differ by $2\,\pi/n\,\mathbf{e}_z$.

To begin understanding the physics of the Hamiltonian~\eqref{eq:hamiltonian_alternative}, we diagonalize it to obtain the band structure shown in Fig.~\ref{fig:dispersion}.
Anticipating the key differences between even and odd $n$, and the central role played by the presence or absence of reflection symmetry along $z$, we show the band structure along the path through momentum space defined in panel~(b) for representatives of the four possible cases in panels (c) to (f).
We find that in all but the reflection-symmetry broken case with odd $n$, the low-energy band structure has a nodal line.
All realizations have low-energy surface states that are degenerate in the presence of reflection symmetry.
To obtain a deeper understanding of these observations, in the next two sections we derive an effective low-energy theory and then provide a symmetry classification of the emergent nodal line.

For our later symmetry analysis, it is important to note that there are $n$ equivalent definitions of the bulk unit cell [an example is shown in Fig.~\ref{fig:sublattice}~(a)].
It will therefore prove useful to define the translation operator
\begin{equation}
 T_{n,m} = \left( \begin{array}{cc} & \mathbb{1}_{m}\\
 \mathbb{1}_{n-m}\,\e^{\im\,n\,k_z} & \end{array} \right) \otimes \sigma_0,
\end{equation}
where $\mathbb{1}_m$ is the an identity matrix of size $m$.
The operator $T_{n,m}$ translates the Hamiltonian's unit cell by $m$ sites and satisfies $T_{n,m} T_{n,n-m} = \e^{\im\,n\,k_z}$.
The transformation
\begin{equation}
 T_{n,m} \,\mathcal{H}_{n\,\mathbf{k}} \,T_{n,m}^\dagger \to \mathcal{H}_{n\,\mathbf{k}}'
\end{equation}
is equivalent to changing all angles of the perturbation \eqref{eq:perturbation} as $\theta_\mu \to \theta_\mu + 2\,\pi\,m/n$.
All bulk properties stay invariant upon such a transformation.

\section{Emergent nodal phases}
\label{sec:low_energy}

To understand the emergence of a nodal line, we introduce a low-energy approximation developed by projecting the full Hamiltonian to bands close to the Weyl nodes.
As described above, the system may be seen as $n$ copies of the Hamiltonian that realize the Weyl phase, Eq.~\eqref{eq:two_band_model}, coupled at different momenta, Eq.~\eqref{eq:hamiltonian_alternative}.
For a folding degree of $n=2$, this especially means that the full Hamiltonian equals two copies of the Weyl Hamiltonian that are coupled at opposite chirality,
\begin{equation}
 \tilde{\mathcal{H}}_{n=2,\mathbf{k}} =
 \begin{pmatrix}
  \mathcal{H}_0 (\mathbf{k} - \pi\,\mathbf{e}_z ) & U_+ + U_- \\
  U_+ + U_- & \mathcal{H}_0 (\mathbf{k} + \pi\,\mathbf{e}_z )
 \end{pmatrix} .
 \label{eq:hamiltonian_n2_alternative}
\end{equation}
This relatively simple $4\times 4$ Hamiltonian allows a detailed investigation of the model's properties in this special case of $n=2$, including an extensive phase diagram.
This is discussed in  the appendix.

For $n>2$, the full Hamiltonian can be expanded around the Weyl nodes at $(0,0,\pm \pi/n)$ and projected down to the lowest bands, giving a form similar to Eq.~\ref{eq:hamiltonian_n2_alternative}, i.e., two Weyl nodes of opposite chirality that are coupled by the potential $U_\pm$.
Up to linear order in momentum and first order\footnote{Orders of $\mathcal{O} (U^2)$ appear on the diagonal of this matrix. They are additionally suppressed by the eigenvalues of $\mathcal{H}_{\mathbf{k} \pm 3\,\mathbf{k}_0}$, i.e., they do not play a role close to the nodal line.} in $U$, the low-energy Hamiltonian has the form
\begin{align}
 \mathcal{H}_\mathbf{k}^\mathrm{low}
	=& v\left( k_x \sigma_x + k_y\sigma_y \right) + v\,\sqrt{1-m^2}\,q_z\,\sigma_z \tau_z \nonumber \\
	& + U_+ \,\tau_+ + U_-\,\tau_-,
\label{eq:actual_low_energy}
\end{align}
where $q_z$ is the momentum along $z$ measured from the Weyl point.
The matrices $\tau_\mu$ act in the space of the different Weyl nodes that are coupled by $U_\pm$.
We introduce the Euclidean Dirac matrices
\begin{subequations}\begin{align}
  \gamma_1 &= \sigma_x,  &  \gamma_2 &= \sigma_y,	\\
  \gamma_3 &= \sigma_z\,\tau_z, & \gamma_4 &= \e^{\im\,\theta_z\,\tau_z}\,\tau_x \sigma_z,
 \end{align}\end{subequations}
and $\gamma_5 = \gamma_1\,\gamma_2\,\gamma_3\,\gamma_4 = -\e^{-\im\,\theta_z\,\tau_z}\,\tau_y\,\sigma_z$ plus the identity and the commutators $\gamma_{ij} = -\frac{\im}{2} [\gamma_i,\gamma_j]$ to rewrite Eq.~\eqref{eq:actual_low_energy} in terms of perturbed Dirac fermions\footnote{The exponential phase dependence in $\gamma_4$ and $\gamma_5$ is equivalent to a rotation of the Hamiltonian.}.
It reads, upon rescaling of momenta
\begin{equation}
 \mathcal{H}_\mathbf{k}^\mathrm{low} = \sum_{i=1}^3 k_i\,\gamma_i + u_z\,\gamma_4 + \mathbf{v} \cdot \mathbf{b}' + \mathbf{w} \cdot \mathbf{p},
\end{equation}
with the definitions\cite{Burkov:2011ek}
\begin{align}
  \mathbf{p} &= \left( \gamma_{14},\gamma_{24},\gamma_{34} \right),~\mathbf{b}'= \left( \gamma_{15},\gamma_{25},\gamma_{35} \right), \\
  \mathbf{w} &= \begin{pmatrix}
                 -u_y\,\cos (\theta_y - \theta_z ) \\ u_x\,\cos (\theta_x - \theta_z) \\ -u_0\,\sin (\theta_0 - \theta_z)
                \end{pmatrix},~\mathbf{v} = \begin{pmatrix}
                 -u_y\,\sin (\theta_y - \theta_z ) \\ u_x\,\sin (\theta_x - \theta_z) \\ u_0\,\cos (\theta_0 - \theta_z) )
                \end{pmatrix} . \nonumber
\end{align}
As discussed in Ref.~\onlinecite{Burkov:2011ek}, a perturbation $u_z$ introduces a mass to the Dirac Hamiltonian and all other terms can lead to Weyl or nodal-line semimetal phases.
However, as shown before in Fig.~\ref{fig:dispersion}~(f), the nodal line that forms is not necessarily stable: it may gap out when taking into account higher orders in momentum and $U$.
In Sec.~\ref{sec:symmetries}, the stability of the nodal line is investigated beyond the low-energy approximation.

Let us focus on a case where the emergence of a nodal line in the spectrum simply follows from the low-energy perspective.
For a perturbation $U_\pm = u_0 \,\e^{\pm \im\,\theta_0} \sigma_0$, we may rotate the low-energy Hamiltonian, Eq.~(\ref{eq:actual_low_energy}) by a unitary transformation
\begin{equation}
 \mathcal{U} = \frac{1}{\sqrt{2}} \left( \sigma_x + \sigma_z \right) \e^{\im\,\phi/2 \,\sigma_z},
\end{equation}
with $\phi$ being the polar angle in the $k_x$-$k_y$ plane.
This transformation gives
\begin{equation}
 \mathcal{U}\,\mathcal{H}_\mathbf{k}^\mathrm{low}\,\mathcal{U}^\dagger
 = v\,q\,\sigma_z + v \sqrt{1-m^2}\,q_z \,\sigma_x\,\tau_z  + u_0\,\e^{\im\,\theta_0\,\tau_z} \,\tau_x,
\end{equation}
with $q = \sqrt{k_x^2 + k_y^2}$.
After applying the canonical transformation
\begin{equation}
 \sigma_\pm \to \sigma_\pm \,\tau_z, \quad  \tau_\pm \to \tau_\pm\,\sigma_z ,
\label{eq:canonical_trafo}
\end{equation}
the Hamiltonian takes the simple form
\begin{equation}
 \mathcal{H}_\mathbf{k}' = \Delta_\pm  \,\sigma_z  - v \sqrt{1-m^2} \,q_z\,\sigma_x,
 \label{eq:nodal_surface}
\end{equation}
where we replaced the operator $\hat{\Delta} = v\,q + u_0\,\e^{\im\,\theta_0\, \tau_z}\,\tau_x$ by its eigenvalues $\Delta_\pm = v\,q \pm u_0$.
The nodal line therefore emerges at $q_z=0, v\,q = u_0$.

\section{Symmetry classification of the nodal line}
\label{sec:symmetries}

\begin{figure}
 \includegraphics[width=\linewidth]{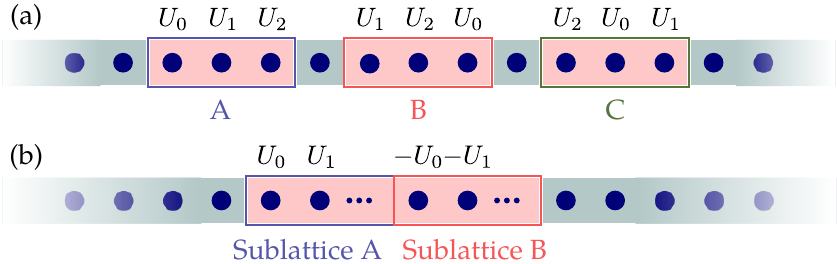}
 \caption{Unit cell in the superlattice. (a) Illustration of the equivalent definitions of the bulk unit cell for a folding degree of $n=3$.
 All bulk properties stay invariant when the definition of the unit cell is changed between the three different options A,B and C.
 (b) For an even folding degree $n$, each unit cell can be split up into two sublattices A and B where the superlattice potential has opposite sign.}
 \label{fig:sublattice}
\end{figure}

\begin{figure}
 \includegraphics[width=\linewidth]{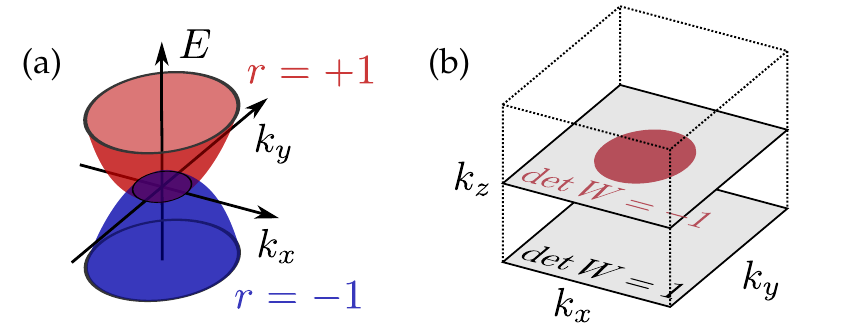}
 \caption{Topological protection of the nodal line.
 The nodal line at $k_z = \pi/n$ is protected by two mechanisms:
 (a) In the presence of reflection symmetry, the two bands that form the nodal line at the reflection invariant momentum $k_z = \pi/n$ have different eigenvalues under reflection, $r= \pm1$.
 The number of occupied bands that are even under reflection changes by $1$ when crossing the nodal line.
 (b) When the Hamiltonian respects the symmetry induced by $C_\mathbf{k}$, the determinant of the matrix that diagonalizes the Hamiltonian, $W (\mathbf{k})$, is quantized to $\det W (\mathbf{k}) = 1$ in the plane defined by $k_z = 0$ and quantized to $\det W (\mathbf{k}) = \pm 1$ in the plane defined by $k_z = \pi/n$, as shown in the figure.
 The red area, within the nodal line, denotes $\det W (\mathbf{k}) = -1$, while the gray area denotes $\det W (\mathbf{k}) = +1$.
 This protection mechanism is only possible for even folding degree $n$.}
 \label{fig:protection}
\end{figure}

Generally, there are two different ways to obtain a protected nodal line in the model \eqref{eq:hamiltonian}: by fulfilling reflection symmetry or a combination of a fractional lattice translation and charge-conjugation symmetry---we explain this in the current section.
The symmetry classification of the nodal line depends on the folding degree of the unit cell $n$.
In its most general form, the full Hamiltonian, Eq.~\eqref{eq:hamiltonian}, does not possess any commuting anti-unitary or anticommuting unitary symmetries, i.e., it is in symmetry class A.
The system is reflection symmetric along $z$ if the operator
\begin{equation}
 \mathcal{R} = 
 \begin{pmatrix}
  & & \sigma_0 \\
  & \iddots & \\
  \sigma_0 & & 
 \end{pmatrix}
\end{equation}
commutes with $\mathcal{H}_{n\,\mathbf{k}}$ or with any shifted Hamiltonian $T_{n,m}\,\mathcal{H}_{n\,\mathbf{k}}\,T_{n,m}^\dagger$ at momenta that are invariant under reflection along $z$, i.e., $k_z = 0$ and $k_z = \pi/n$, where we choose $\mathcal{R}$ such that $\mathcal{R}^2 = +1$, using the convention by Ref.~\onlinecite{Chiu:2014fi}.
This is the case when all angles $\theta_\mu$ are the same and equal $\theta_\mu =  m\,\pi/n$ for all $\mu$ with the integer $m$.
At $k_z = 0$ and $k_z = \pi/n$, the eigenstates of the Hamiltonian are simultaneously eigenstates of $\mathcal{R}$, $\mathcal{R} \ket{\psi} = r \,\ket{\psi}$ with eigenvalues $r= \pm 1$ (the bands are either even or odd under reflection).
Bands with different eigenvalues $r$ cannot mix, which is the mechanism that protects the nodal line, cf.\ Fig.~\ref{fig:protection}~(a).
The invariant characterizing this protection is the mirror Chern number\cite{Chiu:2014fi}: the difference in the number of occupied bands that are even under reflection within and outside the nodal line.

\begin{figure}
 \includegraphics[width=\linewidth]{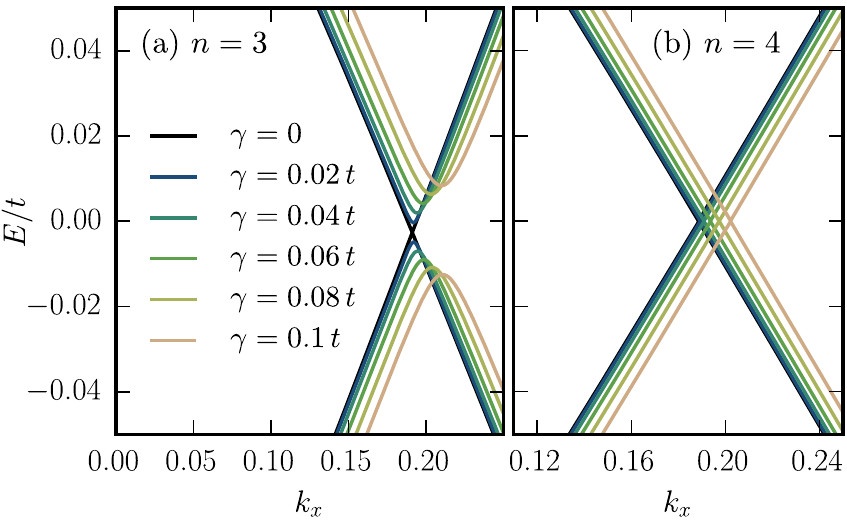}
 \caption{Dispersion of the system close to the nodal line in presence of the reflection-symmetry-breaking term $\gamma\,\sin k_z \,\sigma_0$, with $v=t$, $u_{0} =0.2\,t$, and $u_{x,y,z} = 0$, evaluated at $k_y = 0.05$, $k_z = \pi$.
 (a) For odd folding degree $n$, the reflection-symmetry-breaking term $\gamma$ immediately opens a gap in the spectrum.
 (b) For even folding degree, the gap is stable due to the combination of a fractional lattice translation and charge-conjugation symmetry.}
 \label{fig:finite_gamma}
\end{figure}

For odd folding degree $n$, reflection symmetry is the only symmetry that protects the nodal line.
In absence of reflection symmetry, the nodal line gaps out, giving rise to two Weyl nodes.
This can be achieved in two ways. 
One relies on setting at least one angle to be $\theta_\mu \neq m\,\pi/n$, which opens up a gap along the nodal line, cf.\ Fig.~\ref{fig:dispersion}~(f) for $n=3$.
A second option is to break mirror symmetry by adding a term $\gamma\sin k_z\,\sigma_0$ to the Weyl Hamiltonian, Eq.~(\ref{eq:two_band_model}).
A nonzero $\gamma$ shifts the Weyl nodes to different energies, breaking mirror symmetry and opening up a gap in the nodal line, cf.\ Fig.~\ref{fig:finite_gamma}.

For even $n$, the system is partitioned into two sublattices A and B, cf.\ Fig.~\ref{fig:sublattice}~(b), where the superpotential has opposite sign.
The system can be understood as a one-dimensional system in the $z$-direction, parametrized by the other momentum components $\mathbf{k}_\parallel = (k_x,k_y)$. For a perturbation $U_\pm = u_0 \e^{\pm \im\,\theta_0} \sigma_0$, it obeys a one-dimensional charge-conjugation symmetry
\begin{equation}
 C_\mathbf{k} \,\mathcal{H}_{\mathbf{k}_\parallel,k_z}\,C_\mathbf{k}^{-1} = - \mathcal{H}_{\mathbf{k}_\parallel,-k_z}
 \label{eq:nonsymmorphic}
\end{equation}
where the operator $C_\mathbf{k}$ consists of a combination of a fractional lattice translation and the anti-unitary charge-conjugation symmetry
\begin{equation}
 C_\mathbf{k} = 
 \begin{pmatrix}
  0 & \e^{-\im\,n\,k_z} \\
  1 & 0
 \end{pmatrix} \otimes \sigma_x \mathcal{K},
\end{equation}
with the outer matrix acting in the sublattice space of A, B and $\mathcal{K}$ denoting complex conjugation.
The operator $C_\mathbf{k}$ squares to $C_\mathbf{k}\,C_{-\mathbf{k}} = \e^{-\im\,n\,k_z}$.
Similar to one-dimensional superconductors, the one-dimensional Zak phase
\begin{equation}
 \gamma_j \left( \mathbf{k}_\parallel \right)  = \im\,\int_0^\frac{2 \pi}{n} \di k_z \,\bra{u_{j\,\mathbf{k}}} \partial_{k_z} \ket{u_{j\,\mathbf{k}}}
\end{equation}
is related to the determinant of the matrix $W (\mathbf{k})$ that diagonalizes the Hamiltonian via
\begin{equation}
\exp \left[ \im \sum_{j \in \mathrm{occ.}} \gamma_j \right] = \frac{\det W (\mathbf{k}_\parallel,0)}{\det W(\mathbf{k}_\parallel,\frac{\pi}{n} )} .
\end{equation}
At planes in momentum space defined by $k_z=0$, the operator $C_\mathbf{k}$ squares to $-1$, i.e., the Hamiltonian that respects Eq.~\eqref{eq:nonsymmorphic} is in symmetry class C.
The matrix $W (\mathbf{k}_\parallel,0)$ is symplectic\cite{Altland:1997cg}, which implies that $\det W(\mathbf{k},0) = 1$.
Similarly, at $k_z = \pi/n$, the operator $C_\mathbf{k}$ squares to $+1$, i.e., the Hamiltonian is in symmetry class D.
Hamiltonian matrices in class $D$ can be rewritten in terms of Majorana modes as a skew symmetric matrix $X^T = - X$, which can be diagonalized by an orthogonal matrix\cite{Chiu:2015ex,Budich:2013it}, i.e., $\det W (\mathbf{k}_\parallel,\pi/n) = \pm 1$.
The value of $\det W (\mathbf{k}_\parallel,\pi/n) = \mathrm{sign} \left( \mathrm{Pf} \left[ X_{\mathbf{k}_\parallel} \right] \right)$ is a zero-dimensional invariant that characterizes the two distinct sectors, which are separated by the nodal line.
This invariant is shown in Fig.~\ref{fig:protection}~(b) at both planes $k_z = 0$ and $k_z = \pi/n$.
Its value is $\det W = -1$ inside the nodal line in the plane defined by $k_z = \pi/n$ and $\det W = 1$.
The gap closing cannot vanish without breaking the symmetry protecting this invariant.
Alternatively, the Zak phase can be interpreted as a one-dimensional invariant that is quantized for loops enclosing the nodal line.

Note that the presence of charge-conjugation symmetry does not change the previous statements about the protection of the nodal line in presence of reflection symmetry, since the mirror Chern number is inherited from class A\cite{Chiu:2015ex}.

A particularly simple instance of an even $n$ is the case with folding degree $n=2$. 
In this case the Hamiltonian is always invariant upon a reflection along $z$, i.e.,
\begin{equation}
 \mathcal{R}_{k_z} \,\mathcal{H}_{n=2,\mathbf{k}_\parallel,-k_z} \,\mathcal{R}_{k_z}^\dagger = \mathcal{H}_{n=2,\mathbf{k}_\parallel,k_z}
\end{equation}
with $\mathcal{R}_{k_z} = \cos k_z -\im\,\sin k_z \,\tau_z$.
Analogous to the other reflection-symmetric cases, the bands in the plane defined by $k_z = \pi/2$ have different eigenvalues of $\mathcal{R}_{\pi/2}$, $r=\pm1$, and cannot mix.
The nodal line is therefore protected by a mirror $M \mathbb{Z}$ Chern number.
Additionally, the presence of charge-conjugation symmetry protects the nodal line even if reflection symmetry is broken by a term $\gamma\,\sin k_z\,\sigma_0$, a fact holds for all systems with even folding degree $n$.

\section{Surface states}
\label{sec:surface_states}
In the model investigated here, surface states play a crucial role: it is possible to have realizations of a stable nodal-line phase with drumhead surface states as well as surface states that are not exponentially localized at the boundary.
The emergence of surface states close to zero energy is discussed in this section.

\begin{figure}
 \includegraphics[width=\linewidth]{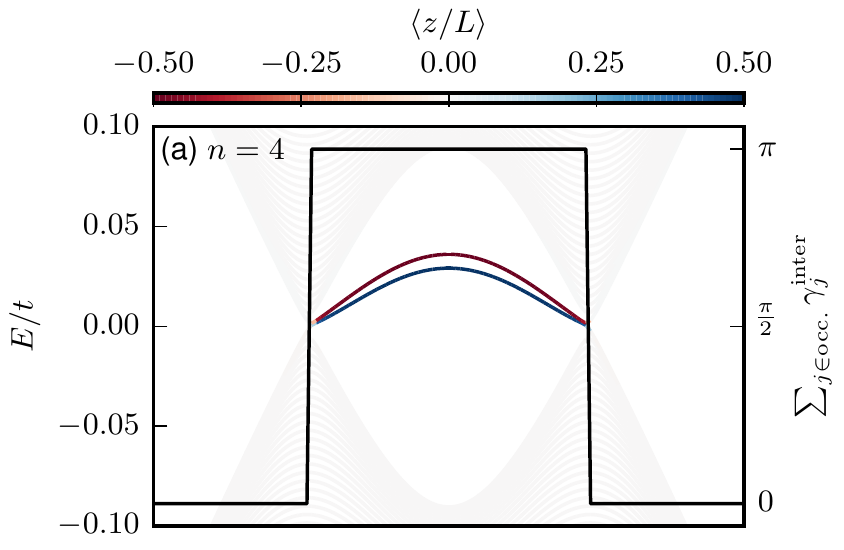}
 \includegraphics[width=\linewidth]{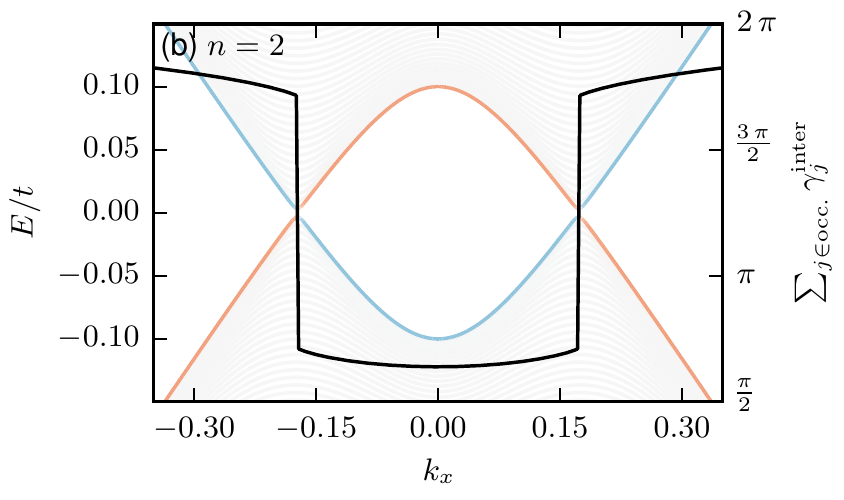}
 \caption{Energy dispersion and intercellular Zak phase of a periodically perturbed Weyl semimetal that is finite in $z$ with $L = 512$ sites, $-L/2<z<L/2$, evaluated at $k_y = 0.05$.
 The color of the lines encodes the average real-space position.
 (a)~Folding degree $n=4$ and perturbation strength $u_0 = 0.2\,t$ with $\theta_0 = 3/4\,\pi$.
 Intercellular Zak phase and total Zak phase coincide and are both quantized to $0$ (outside the nodal line) and $\pi$ (within the nodal line).
 The intercellular Zak phase predicts the extra-charge accumulation at the surface, which results in exponentially localized surface states (the degeneracy of the surface states is lifted by an extra energy $\eta = 0.005\,t$).
 (b)~Folding degree $n=4$ and perturbation strength $u_0 = 0.2\,t$ with $\theta_0 = 0$.
 Although the total Zak phase is quantized, the intercellular Zak phase is not; this results in surface states that are not exponentially localized at the surface.
 Furthermore, the surface states are not locked to small energies.}
 \label{fig:surface_charge}
\end{figure}

In Fig.~\ref{fig:surface_charge}~(a), the energy dispersion of a finite system with folding degree $n=4$ that respects reflection symmetry is plotted, along with the location of the states encoded by a color scale.
The system exhibits drumhead surface states at small energies that are exponentially localized at the surface.
However, not all realizations of a stable nodal line share this feature:
an example without drumhead surface states is given in Fig.~\ref{fig:surface_charge}~(b), where the dispersion for a finite system with folding degree $n=2$ is shown.
It is not possible to find a surface termination that respects the reflection symmetry of the bulk system.
This results in states that are not exponentially localized at the surfaces of the system.	
To explain the properties of the surface states, we introduce and compute the Zak phase.

In the modern theory of polarization\cite{Vanderbilt:1993dx}, the Zak phase is associated with the surface charge.
Here, we employ the intercellular Zak phase that reflects the choice of a bulk unit cell to relate the extra charge accumulation at the surface with bulk properties of the system\cite{Rhim:2017bg}.
The intercellular Zak phase and extra charge accumulation are proportional when the finite system is commensurate with the unit cell used to compute the intercellular Zak phase.
While the Zak phase is generally given by
\begin{equation}
 \gamma_j (k_x,k_y)= \im\,\int_0^{2\,\pi/n} \di k_z \bra{ u_{j\,\mathbf{k}} } \partial_{k_z} \ket{ u_{j\,\mathbf{k}}}
\end{equation}
with the lattice-periodic part of the wave function $u_{j\,\mathbf{k}} (\mathbf{r}) = \e^{-\im\,\mathbf{k} \cdot \mathbf{r}} \,\psi_{j\,\mathbf{k}} (\mathbf{r})$, we rather focus on
\begin{equation}
 \gamma_j^\mathrm{inter} (k_x,k_y ) = \im\,\int_{0}^{2\,\pi/n} \di k_z \, \bra{ \psi_{j\,\mathbf{k}} } \partial_{k_z} \ket{ \psi_{j\,\mathbf{k}}} ,
\end{equation}
neglecting the ``classical'' contribution to the surface charge, coming from the polarization within a unit cell.
The functions $\ket{\psi_{j\,\mathbf{k}}}$ are chosen such that all sites in one unit cell carry the same Bloch phase\cite{Rhim:2017bg}.
This quantity can be numerically evaluated efficiently by calculating the corresponding Wilson loop\cite{Alexandradinata:2014ju}.

The intercellular Zak phase is only quantized to $0,\pi$ for a reflection-symmetric unit cell, when it equals the total Zak phase, otherwise it can take an arbitrary value.
Extra charge accumulation $Q^{L(R)}_\mathrm{acc.}$ and intercellular Zak phase are related via\cite{Rhim:2017bg}
\begin{equation}
 Q^{L(R)}_\mathrm{acc.} = +(-) \frac{e}{2\,\pi}\sum_{j\in \mathrm{occ.}} \gamma_j^\mathrm{inter} (\mathrm{mod}~e)
 \label{eq:extra_charge}
\end{equation}
where $L(R)$ stands for the left (right) surface region commensurate with the bulk unit cell used for the calculation of the intercellular Zak phase.
Furthermore, Eq.~\eqref{eq:extra_charge} does not allow us to make a statement about the energy of the extra-charge accumulation.
It has been shown that the above relation leads to the $\mathbf{Z}_2$ bulk-boundary correspondence for the reflection symmetric insulators: we have $\gamma^\mathrm{inter}/\pi$ (mod 2) in-gap surface modes, with $\gamma^\mathrm{inter} = \sum_{j\in \mathrm{occ.}} \gamma_j^\mathrm{inter}$, if the finite system (i) respects the reflection symmetry and (ii) is commensurate with the bulk unit cell\cite{Rhim:2017bg}.

As shown in Fig.~\ref{fig:surface_charge}~(a), the intercellular Zak phase for a reflection symmetric system with folding degree $n=4$ is quantized to $0,\pi$, correctly predicting the number of in-gap surface modes.
However, the $n=2$ case violates the condition (i) for the bulk-boundary correspondence, and surface modes are not guaranteed to exist although the total Zak phase is quantized.
Instead, the intercellular Zak phase explains the extra charge accumulations for both surfaces correctly as plotted in Fig.~\ref{fig:surface_charge}~(b).

\section{Stability against wave vector mismatch}
\label{sec:stability}
In the previous discussion, the focus was on a perfect match of the superlattice's wave vector and the node separation of the underlying Weyl semimetal.
In any realistic system, such a perfect agreement may be hard to realize.
There are two possibilities for a vector mismatch: we can change the superlattice's wave vector $\mathbf{K} = \left( 2\,\pi/n  + \delta k \right) \mathbf{e}_z$ or we can change the node separation by changing $m \to \eta/v + \cos \pi/n$.

\begin{figure}
 \includegraphics[width=\linewidth]{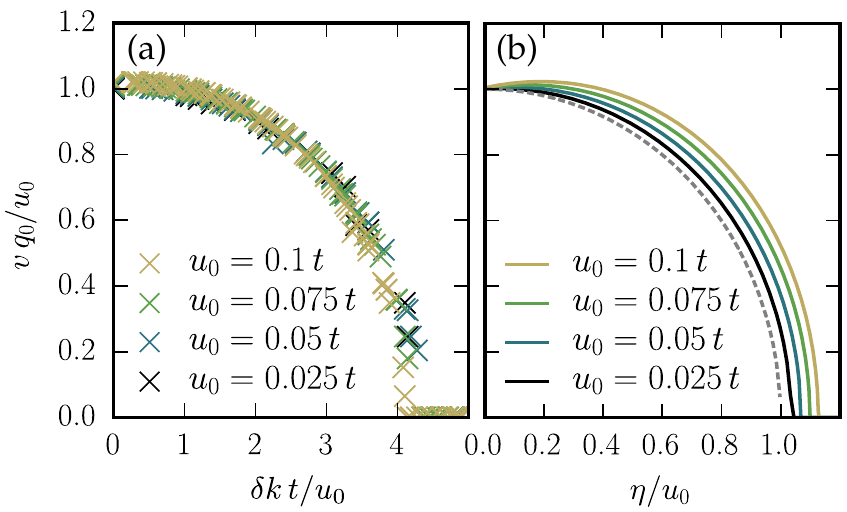}
 \caption{Stability of nodal line.
 Position of nodal line $(q_0,0,\pi/n)$ for a periodically perturbed Weyl semimetal with a superlattice of folding degree $n=7$ for different potential strengths $u_0$, with $v=t$ and $u_{x,y,z} = 0$.
 (a) The superlattice is varied $\mathbf{K} \to \left( 2\pi/n + \delta k\right) \,\mathbf{e}_z$ so that is does not match the node separation.
 (b) The mass parameter is varied $m \to \cos \pi/n + \eta/v$ so that the node separation does not match the superlattice.
 Solid lines and crosses show the results using the full Hamiltonian, and the dashed line shows the lowest order prediction.}
 \label{fig:stability}
\end{figure}

When the wave vector $\mathbf{K}$ is changed, it generally fits neither the node separation nor the lattice.
In such a case, translational invariance is lost, as investigated in Ref.~\onlinecite{Wang:2017ib}.
Here, we are interested in the fate of the nodal line in presence of a wave vector that does not match the node separation, i.e., we choose a vector $\mathbf{K}$ that is commensurate with periodic boundary conditions.
In Fig.~\ref{fig:stability}, the radius of the nodal line is shown for several lattices with periodic boundary conditions that allow certain $\mathbf{K}$.
The diameter of the nodal line shrinks with increasing $\delta k$ until it vanishes at a critical value that depends on $u_0$ and the folding degree $n$.

By changing $m \to \eta/v + \cos \pi/n$, the position of the Weyl nodes in the original unperturbed system are modified.
The position of the nodal line can be obtained from modifying Eq.~(\ref{eq:nodal_surface}).
This gives the condition for the position in the $x$-$y$-plane with $q=\sqrt{k_x^2 + k_y^2}$
\begin{equation}
 v\,q = \sqrt{u_0^2 - \eta^2} ,
\end{equation}
i.e., we expect that the nodal line stays stable up to $\left| \eta \right| \le \left|u_0 \right|$.
In Fig.~\ref{fig:stability}, the lowest-order prediction is compared with numerical results for the full Hamiltonian.
As in the previous case, the diameter of the nodal line shrinks for nonzero $\eta$, but it generally persists up to a critical value, in agreement with the symmetry classification that generally predicts the stability of the nodal line and does not rely on specific values of the mass term.

\section{Time-reversal symmetric Weyl semimetal}
\label{sec:tr_symmetric}

\begin{figure}
 \includegraphics{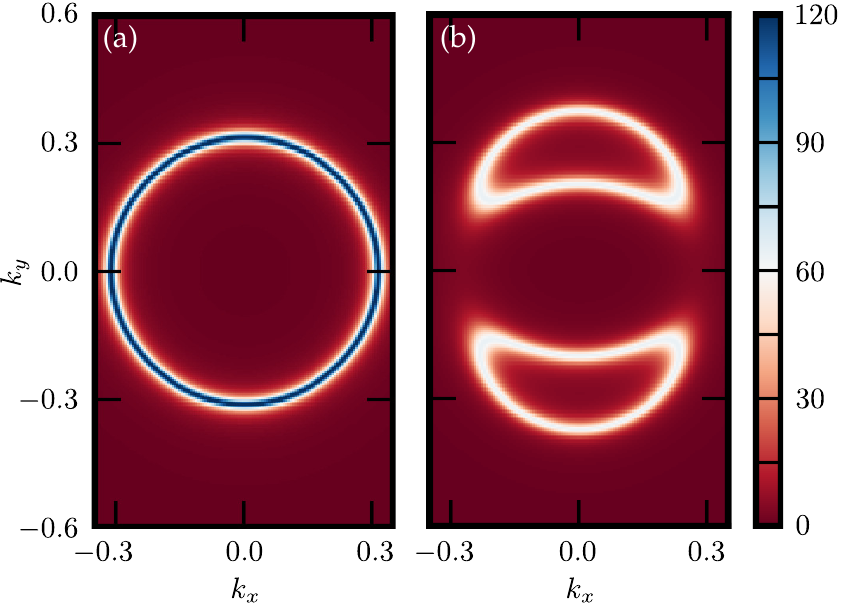}
 \caption{Density of states at zero energy for the time-reversal invariant system subjected to a periodic perturbation $U (\mathbf{r} ) = 2\,u_{00}\,\cos \left( 2\,\mathbf{r} \cdot \mathbf{K} -\theta_{00} \right)\,s_0\,\sigma_0$ with folding degree $n=4$ and $v = t, u_{00} = 0.3\,t$, $\theta_{00} = 3/4\,\pi$.
 The level broadening $\eta$ is chosen $\eta = 0.01\,t$.
 (a) Starting from the Dirac phase with $B_1 = \mathbf{k}_1 =0$, a fourfold-degenerate nodal line emergence.
 Splitting up the original Dirac node into two Weyl nodes with $B_1= 0.1\,t$ and $\mathbf{k}_1 = (0,0.15,0)$ lifts one degeneracy and splits the fourfold-degenerate nodal line into two twofold-degenerate nodal lines.}
 \label{fig:fermi_surface_TRI}
\end{figure}

The previous findings can be extended to a model that is closer to currently available materials that respect time-reversal symmetry.
Two copies of the Weyl Hamiltonian
\begin{equation}
 \mathcal{H}_0^{(\pm)} (\mathbf{k} ) = v\,\left( \pm \sin k_x \,\sigma_x + \sin k_y \,\sigma_y \right) + M_\mathbf{k}\,\sigma_z
\end{equation}
with $M_\mathbf{k}$ from Eq.~(\ref{eq:two_band_model}) give rise to several phases, including Dirac semimetals, Weyl semimetals with four Weyl nodes and strong topological insulators, all described by\cite{Wang:2012ds}
\begin{equation}
 \mathcal{H}_\mathbf{k} = \left( \begin{array}{cc} \mathcal{H}_0^{(+)} (\mathbf{k} - \mathbf{k}_1) & B_1\,\e^{\im\,\phi}\,\sin(k_z)\,\sigma_x \\
 B_1\,\e^{-\im\,\phi}\,\sin (k_z) \,\sigma_x & \mathcal{H}_0^{(-)} (\mathbf{k} + \mathbf{k}_1)
 \end{array} \right) .
 \label{eq:inv_weyl}
\end{equation}
The vector $\mathbf{k}_1$ lies in the $k_x$-$k_y$-plane and it breaks inversion symmetry.
The outer matrix structure is described by the Pauli matrices $s_\mu$ acting in spin space.
This Hamiltonian is time-reversal symmetric with $\Theta = \im\,s_y\,\mathcal{K}$ and has reflection symmetry along $z$ with $\mathcal{R} = s_z$.
When $\mathbf{k}_1 = 0$ and $B_1 = 0$, the system has a $C_4$ rotational symmetry in the $k_x$-$k_y$-plane.
Then, there are two degenerate Dirac nodes at $(0,0,\pm \arccos m)$.
In absence of inversion and rotational symmetry, the Dirac nodes either split up into four Weyl nodes for $|\mathbf{k}_1| > B_1$ or gap out when $|\mathbf{k}_1| < B_1$.

A periodic perturbation with a wave vector $\mathbf{K} = 2\,\arccos m\,\mathbf{e}_z$ can lead to protected line nodes.
The most general perturbation 
\begin{equation}
 U (\mathbf{r}) = \sum_{\mu,\nu} u_{\mu\nu} \,\cos \left( \mathbf{r} \cdot \mathbf{K} - \theta_{\mu\nu} \right)\,\sigma_\mu\,s_\nu
\end{equation}
may give rise to a plethora of different phases.
For simplicity, we just discuss the simplest case of a onsite potential, i.e., just $u_{00}$ is nonzero (see Fig.~\ref{fig:fermi_surface_TRI}).
In the Dirac phase, such a perturbation can give rise to a fourfold-degenerate nodal line.
In presence of reflection symmetry, the symmetry class is AII with $R_-$, i.e., $\mathcal{R}$ and $\Theta$ anti-commute\cite{Chiu:2014fi}.
Although this symmetry class only allows a $\mathbb{Z}_2$ classification that does not protect the nodal line, the $M\,\mathbb{Z}$ classification inherited from A remains\cite{Chiu:2015ex}.

Breaking inversion and $C_4$ symmetry by $\mathbf{k}_1$ and $B_1$ splits up the fourfold-degenerate nodal line into two twofold-degenerate nodal lines.
The protection mechanism of these nodal lines is analogous to the previously considered time-reversal-breaking case, since the additional time-reversal symmetry does not change the classification.
The operator
\begin{equation}
 \bar{C}_\mathbf{k} = 
 \begin{pmatrix}
    & \e^{-\im\,n\,k_z} \\
   1 &
 \end{pmatrix} \otimes \left( \sin \phi \,s_0 -\im\, \cos \phi\,s_z \right)\sigma_y \mathcal{K}
\end{equation}
anti-commutes with the Hamiltonian and squares to $C_\mathbf{k}\,C_{-\mathbf{k}} = -\e^{-\im\,n\,k_z}$, giving a $\mathbb{Z}_2$ invariant defined in the $k_z = \pi/n$ plane, analogous to the classification for the time-reversal breaking case in Sec.~\ref{sec:symmetries}.

\section{Summary and conclusion}

We showed that a Weyl semimetals subjected to a periodic modulation of the onsite potential can give rise to a nodal-line semimetal that is not gapped out by spin-orbit coupling.
Since the nodal line is protected by mirror symmetry and/or the combination of a fractional lattice translation and charge-conjugation symmetry, its presence does not rely on details of the Hamiltonian.
Although this work focused, because of its simplicity, on a Weyl semimetal with two Weyl nodes as a starting point, we further show that nodal lines also arise in periodically perturbed Weyl semimetals that respect time-reversal symmetry, by using a model that has four Weyl nodes without the perturbation.

An unusual feature of this proposal are the surface states that are not necessarily exponentially localized at the surface close to zero energy;
this may open possibilities for experimental investigations of bulk properties that are not disturbed by any low-energy surface states.
Similarly, quasiparticle interference\cite{Lodge2017} is a promising tool for probing the nodal line at the Fermi level.

\acknowledgments
We thank Robert-Jan Slager, Fernando de Juan, and Takahiro Morimoto for useful discussions. This work was supported by the ERC Starting Grant No.~679722.
A. G. G. was supported by the Marie Curie Programme under EC Grant Agreement No.~653846.

\appendix

\section{Exact Hamiltonian and Phases for a Commensurate Perturbation with Folding Degree $n=2$}
\label{appendix:n2}

\begin{figure*}
\includegraphics[width=2.07\columnwidth]{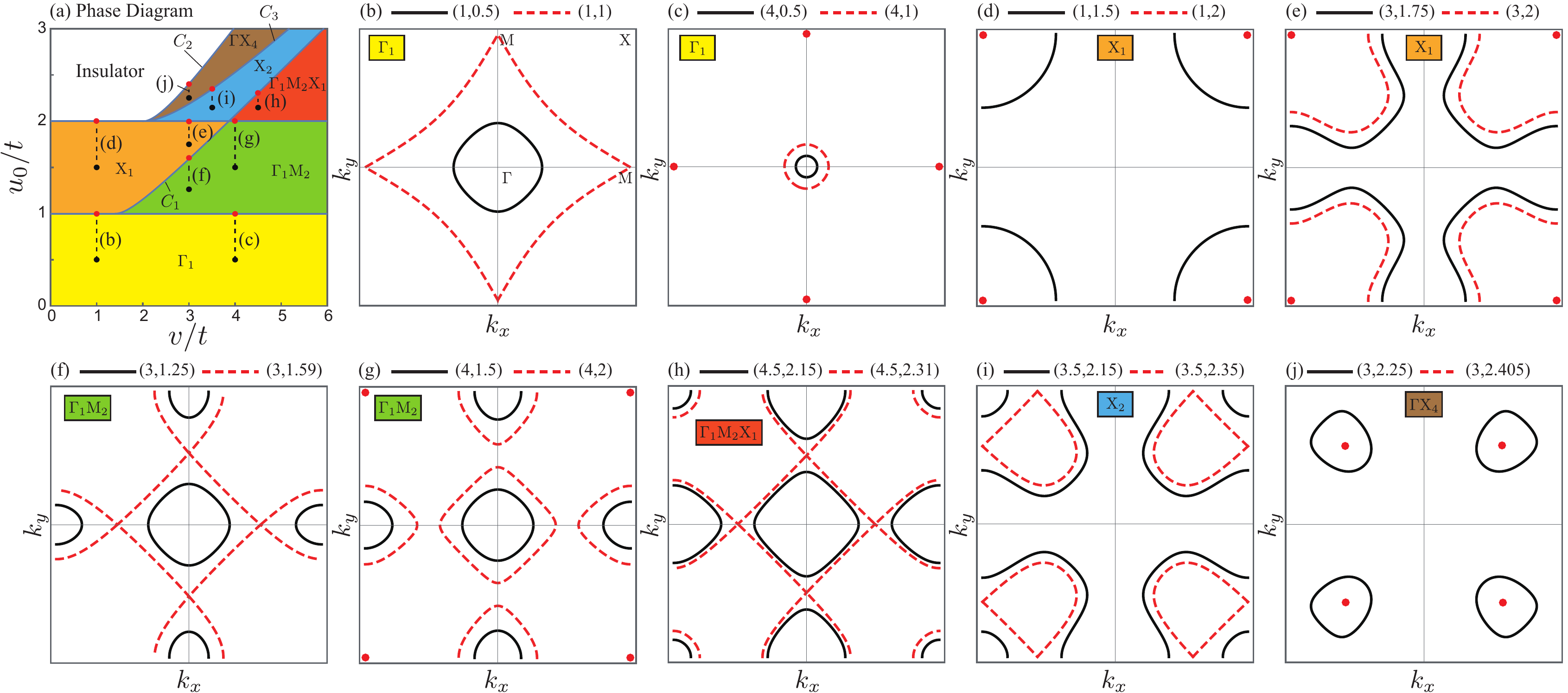}
\caption{(a) Phase diagram for various nodal semimetal phases as a function of $v/t$ and $u_0/t$.
There are six nodal-line phases which are distinguished by different colors.
We classify them based on how many nodal rings are in the Brillouin zone, and where the centers of nodal rings are located.
Those properties of each nodal-line phase are reflected in its name where the capital letters denotes the positions of the centers of nodal rings, and their subscript represent the number of nodal rings centered on them.
As an example, in the phase $\Gamma_1$M$_2$X$_1$, we have one nodal ring around the $\Gamma$ point, two around two M points, and one around an X point as shown in panel (h) by black solid curves.
On the other hand, in the phase $\Gamma$X$_4$, there are four nodal rings around four points on four high symmetry lines $\Gamma$X, as plotted in panel (j) by black solid curves.
At boundaries between those six nodal-line phases, we have intermediate phases which contain singular nodes in addition to nodal lines.
In panels (b) to (j), we plot nodal Fermi surface structures corresponding to the phases marked by black and red dots in panel (a).
On top of each panel, we specify the tight-binding parameters $(v/t,u_0/t)$ for the black solid and red dashed curves.
For example, the band structure for the black dot at $(v/t,u_0/t) = (4,0.5)$ in panel (a) is drawn by the black solid curve in panel (c) whereas the band structure for the red dot at $(4,1)$ in panel (a) is plotted by the red dashed curve and red dots in panel (c).
}
\label{fig:nodal_phase}
\end{figure*}

When the unit cell contains two sites ($n=2$), we can solve the Hamiltonian,
\begin{equation}
 \mathcal{H}_{n=2,\mathbf{k}} =
 \begin{pmatrix}
  h_\mathbf{k} + U_0 & v\,\e^{-\im\,k_z} \,\cos k_z\,\sigma_z \\
  v\,\e^{\im\,k_z} \,\cos k_z\,\sigma_z & h_\mathbf{k} - U_0
 \end{pmatrix}
 \label{eq:n_equals_two}
\end{equation}
fully analytically, and obtain a complete phase diagram as shown in Fig.~\ref{fig:nodal_phase}(a).
From Fig.~\ref{fig:nodal_phase}(b) to (j), we exhibit various evolutions of nodal lines and show how one can achieve the phase transitions between different nodal-line phases through the singular nodes.

When $U (\mathbf{r} ) = 2\,u_0 \cos \left( \pi\, z \right)\sigma_0$, the four eigenvalues of Eq.~\eqref{eq:n_equals_two} are given by
\begin{align}
E^{\eta_1,\eta_2}_{\mathbf{k}} =& \eta_1 \Big[ \left( M_\mathbf{k} - a_\mathbf{k} \right)^2 + 4u_0^2 + a_\mathbf{k}^2 + v_\mathbf{k}^2 \\
& +2\eta_2 \left\{ \left( M_\mathbf{k} - a_\mathbf{k} \right)^2\left( 4u_0^2 + a_\mathbf{k}^2 \right) + 4u_0^2v_\mathbf{k}^2 \right\}^{\frac{1}{2}} \Big]^{\frac{1}{2}} \nonumber\label{eq:eigenvalue_2}
\end{align}
where $a_\mathbf{k} = v\cos k_z$, $v_\mathbf{k} = v(\sin^2k_x + \sin^2k_y)^{1/2}$, and $\eta_i = \pm$.
We assume, without loss of generality, that $v$ is positive.

Because $\left( M_\mathbf{k} - a_\mathbf{k} \right)^2 + 4u_0^2 + a_\mathbf{k}^2 + v_\mathbf{k}^2$ is positive, nodal points can be found from the zeros of $E^{\pm,-}_{\mathbf{k}}$, which are given by
\begin{equation}
a_\mathbf{k}^2 = (M_\mathbf{k} - a_\mathbf{k})^2 - v_\mathbf{k}^2 - 4u_0^2 \pm 2 v_\mathbf{k} \sqrt{ 4u_0^2 - (M_\mathbf{k} - a_\mathbf{k})^2 }.\label{eq:nodal_equation}
\end{equation}
For the right-hand side of (\ref{eq:nodal_equation}) to be real, we must have $4u_0^2 > (M_\mathbf{k} - a_\mathbf{k})^2$ and take the plus sign.
From (\ref{eq:nodal_equation}), we observe that the nodal line can only exist at $k_z = \pi/2$.
In this case, the nodal line is given by
\begin{align}
4u_0^2 = t^2\left( 2- \cos k_x - \cos k_y\right)^2 + v^2\left( \sin^2k_x + \sin^2k_y \right) .\label{eq:nodal_ring}
\end{align}
In the limit of the small external potential ($u_0 \ll 1$), this equation reduces to $ 8u_0^2/(t^2+2v^2) = k_x^2 + k_y^2$ which is a nodal ring enclosing the origin ($\Gamma$ point).
This is consistent with the phase diagram Fig.~\ref{fig:nodal_phase}(a).

Figure~\ref{fig:nodal_phase}(a) shows the full phase diagram of 15 nodal semimetal phases as a function of the parameters $v/t$ and $u_0/t$, obtained  from (\ref{eq:nodal_ring}).
There are six nodal-line semimetal phases denoted by $\Gamma_1$, X$_1$, $\Gamma_1$M$_2$, $\Gamma_1$M$_2$X$_1$, X$_2$, and $\Gamma$X$_4$.
They consist of closed curves as Fermi surfaces, where each of them encloses a certain high symmetry point, or a point on the straight line connecting two high symmetry points.
The name of each nodal-line phase in the above is composed of those high symmetry points, and its subscripts represent the number of nodal rings enclosing them in the Brillouin zone.
For a sufficiently large value of $u_0$, the system becomes insulating.
At the boundaries between those six nodal-line phases and the insulating phase, one finds nine singular nodal phases where nodal contours are not differentiable at several points due to the existence of the nodal point or the crossing between nodal lines.
Those singular points are inevitable to have a phase transition between different nodal-line phases.
The phase boundaries are represented by the curves $u_0/t=1$, $u_0/t=2$, and $u_0/t=C_i(v/t)$ ($i=1,2,3$).
Here, $C_1(x) = x^2/2(x^2-1)^{1/2}$, $C_2(x) = x^2/(2x^2-4)^{1/2}$, and $C_3(x) = x(x^2+8)^{1/2}/(4x^2-4)^{1/2}$.
Those phase transitions are exhibited in Fig.~\ref{fig:nodal_phase}(a) by vertical dashed lines.
The Fermi surfaces corresponding to their lower ends (black dot) and upper ends (red dot) are plotted from Fig.~\ref{fig:nodal_phase}~(b) to (j) by black and red colors.
One can see that the Fermi surfaces of the intermediate singular nodal phases drawn by red dashed curves and red dots always have singular points.

\bibliography{nlsm}

\end{document}